\documentclass[aps,pre,twocolumn,superscriptaddress,floatfix]{revtex4}
\usepackage{graphicx}
\usepackage[dvips,unicode,colorlinks,linkcolor=blue,citecolor=blue,urlcolor=blue]{hyperref}
\usepackage{times}
\begin{document}
\title{Transport of fullerene molecules along graphene nanoribbons}

\author{Alexander V. Savin}

\affiliation{Nonlinear Physics Center, Research School of Physics and
Engineering, Australian National University, Canberra ACT 0200,
Australia}

\affiliation{Semenov Institute of Chemical Physics, Russian Academy
of Sciences, Moscow 119991, Russia}

\author{Yuri S. Kivshar}
\affiliation{Nonlinear Physics Center, Research School of Physics and
Engineering, Australian National University, Canberra ACT 0200,
Australia}

\begin{abstract}
We study the motion of C$_{60}$ fullerene molecules (buckyballs) and short-length carbon nanotubes
on graphene nanoribbons. We demonstrate that the nanoribbon edge creates an effective potential that
keeps the carbon structures on the surface. We reveal that the
character of the motion of C$_{60}$ molecules depends on temperature: for low temperatures ($T<150$~K)
the main type of motion is sliding along the surface, but for higher temperatures the sliding is replaced
by rocking and rolling. Modeling of the buckyball with an included metal ion, such as Fe$^-$@C$_{60}$,
demonstrates that this molecular complex undergoes a rolling motion along the nanoribbon
with the constant velocity under the action of a constant electric field.
The similar effect is observed in the presence of the heat gradient applied to the nanoribbon,
but mobility of carbon structures in this case depends largely on their size and symmetry,
such that larger and more asymmetric structures demonstrate much lower mobility. Our results suggest
that both electorphoresis and thermophoresis can be employed to control the motion of carbon molecules
and fullerenes and, for example, sort them by their size, shape, and possible inclusions.
\end{abstract}

\maketitle

\section{Introduction}

Over the past 20 years nanotechnology has made an impressive impact on the development of many fields
of physics, chemistry, medicine, and nanoscale engineering~\cite{book}. After the discovery of graphene
as a novel material for nanotechnology~\cite{science}, many properties of this two-dimensional object
have been studied both theoretically and experimentally~\cite{review,review2}. In addition to its unusual
properties, graphene may play an important role as a component of more complex systems of nanotechnology.
In particular, atoms or molecules deposited on a surface of single-layer graphene~\cite{nature} 
are expected to move with just two degrees of freedom, demonstrating various phenomena that can be employed
for nanomechanics and nanoscale molecular motors~\cite{science2}.

Numerical modeling of the motion of atoms and molecules on a graphene sheet was considered very recently
in a number of studies~\cite{nl10cms,narne10pre,iams10jp}. In the first approximation, one may treat
graphene as a planar two-dimensional substrate that creates an effective
two-dimensional periodic potential for absorbed molecules. Stochastic motion of noble gases on
such a substrate potential was analyzed in Ref.~\cite{nl10cms}.  It was shown
that Xe atom is trapped in a potential well of the substrate at high friction while
He atom can freely diffuse. Diffusive motion of C$_{60}$ molecules on a graphene sheet was modeled
in Ref.~\cite{narne10pre}. Analytical and numerical studies demonstrated that the translation
motion of C$_{60}$ molecule near a graphene sheet is diffusive in the lateral direction, whereas
in the perpendicular direction this motion can be described as diffusion in an effective harmonic
potential (bounded diffusion). It was also shown that the motion of C$_{60}$ over
graphene sheet is not rolling. More detailed analysis of the structure
of the effective two-dimensional periodic potential for C$_{60}$ molecules
was carried out in Ref.~\cite{iams10jp}. It was shown that the energetically favorable motion
of fullerene molecules is sliding along the surface when one side of the C$_{60}$
molecule remains parallel to the plane of the graphene and the molecule center is placed
above the valent bond of the nanoribbon. However, all those results were obtained for the case
when the graphene nanoribbon is assumed to be flat and rigid, so that the motion of its atoms
is not taken into account.

Directed motion of C$_{60}$ molecules along the graphene sheet under the action of a heat flow
was studied numerically in Ref.~\cite{lne11pre}. In their calculations, the authors employed
the so-called Nos\'e-Hoover thermostat, which however is known to produce incorrect results
in the nonequilibrium molecular dynamics simulations \cite{gss04,llm09}. 
In addition, the directed motion of the fullerenes were not analyzed.

By now, novel nanocomposite structures consisting of a stacked single graphene sheet and  C$_{60}$
monolayer were produced experimentally~\cite{ikysksm10nanomat}.  It was found that the
intercalated C$_{60}$ molecules can rotate in between graphene sheets. Such a rotation of a single
C$_{60}$ molecule on a sheet of graphene should result into its rocking and rolling motion,
in contrast to the statement of Ref.~\cite{narne10pre}. In order to resolve those controversies
and model more general dynamics closely resembling the experimental conditions, 
one have to take into account
the motion of carbon atoms of the nanoribbon that will modify its planar geometry due
to the interaction with C$_{60}$ fullerene molecules.

In this paper, we study numerically different types of motion of carbon structures, such as C$_{60}$
fullerene molecules and short-length carbon nanotubes, on finite-width graphene nanoribbons.
We employ molecular dynamics simulations taking into account the motion of all carbon atoms, and demonstrate
that an edge of the nanoribbon creates a potential barrier for the absorbed nanostructures supporting
their transport along the nanoribbon. We study different types of dynamics including stochastic Brownian
like motion as well as directed motion of fullerenes driven by the electric field and temperature gradient.
We reveal that the motion of C$_{60}$ fullerene molecules depends on the temperature, so that
for low temperatures the main type of motion is sliding along the surface, but for higher temperatures
the sliding is replaced by rocking and rolling. For the directed motion,  the mobility depends largely
on the shape and symmetry of the carbon structures, so that larger and asymmetric structures demonstrate
much lower mobility. Our results suggest that both electorphoresis and thermophoresis can be employed
to control and direct carbon structures and fullerenes.
\begin{figure}[tbp]
\begin{center}
\includegraphics[angle=0, width=0.9\linewidth]{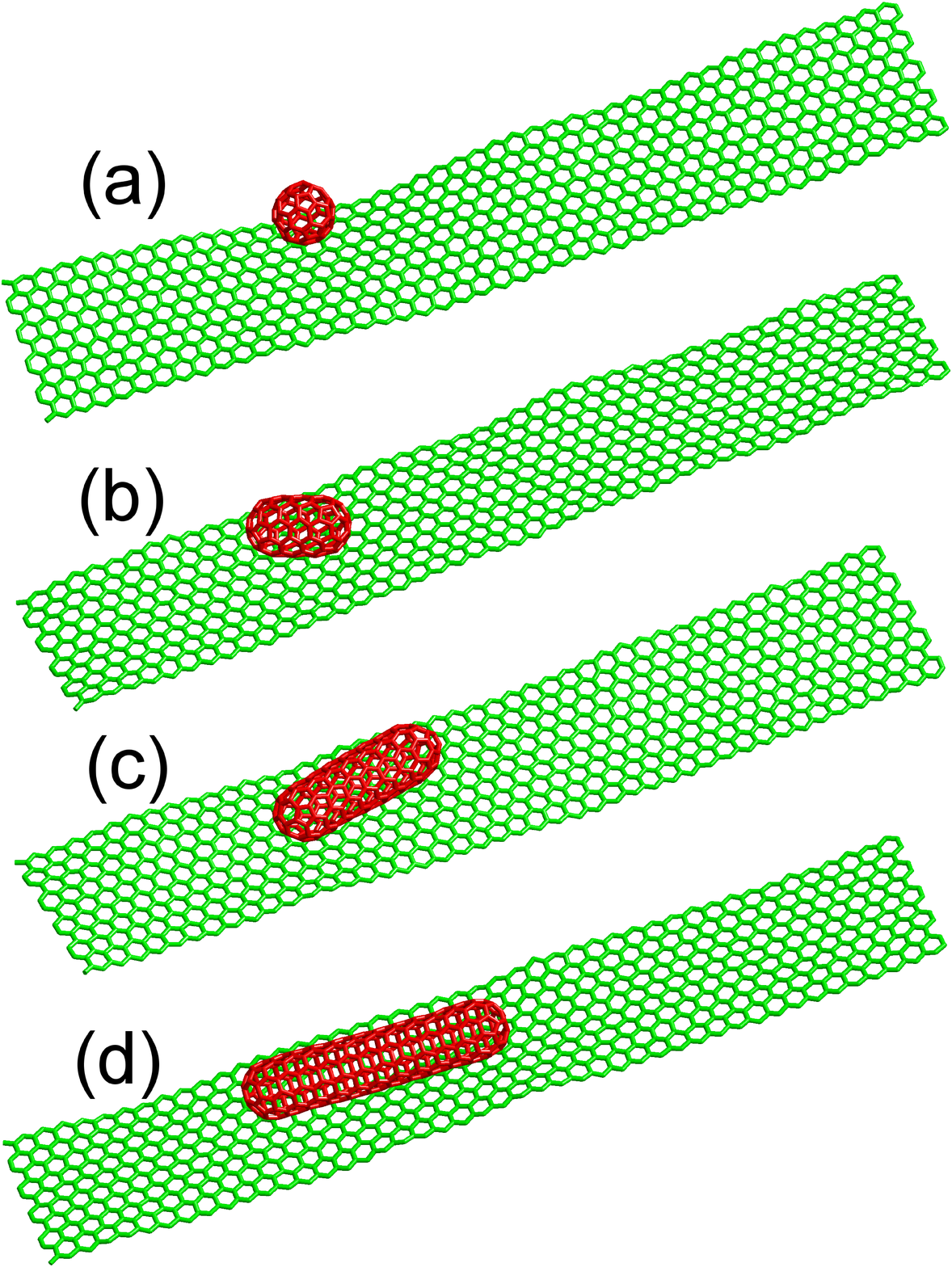}
\end{center}
\caption{\label{fig01}\protect
Motion of (a) fullerene C$_{60}$ molecule, and (b-d) shot-length (5,5) carbon nanotubes
 C$_{100}$, (c) C$_{180}$, and (d) C$_{260}$, respectively, along a finite-width
zigzag carbon nanoribbon. Valent bonds of the nanoribbon are shown in green, while the
valent bonds of the absorbed fullerene-like molecules, are shown in red. Temperature is
 $T=300$K.
}
\end{figure}

\section{Model}

The first study of the van der Waals (VDW) energy for the interaction of fullerene molecule C$_{60}$
and substrate with atomic surface corrugation was presented in Ref.~\cite{gdlbgl96prb}. The VDW energy
was calculated using linear response theory to evaluate the dipole-dipole interactions between the molecule
and the substrate. Interaction of the fullerene C$_{60}$ with graphite surface was described by
the 12-6 Lennard-Jones (LJ) potential
\begin{equation}
V(r)=C_{12}/r^{12}-C_{6}/r^6 \label{f1}
\end{equation}
with the coefficients $C_{12}=12000$~eV\AA$^{12}$ and $C_6=15.2$~eV\AA$^6$.

Experimental evaluation of the interaction of molecule C$_{60}$
with single-wall carbon nanotubes (SWNTs) and graphite was obtained in Ref.~\cite{umh03prl},
on the basis of the thermal desorption spectroscopy. It was shown that these interactions
can be modeled, with a high accuracy, by the LJ potential (\ref{f1}) with
coefficients  $C_{12}=22500$~eV\AA$^{12}$ and $C_6=15.4$~eV\AA$^6$. This potential has
been employed for the modeling absorption of C$_{60}$ molecules by single-walled
carbon nanotubes with open ends.

Detailed modeling of the motion of a C$_{60}$ molecule inside a single-walled
carbon nanotube was carried out in a series of studies~\cite{xc06nt,sgz06nt,hx09pla,rh10cpl}.
In particular, Ref.~\cite{xc06nt} studied the motion of C$_{60}$ inside (11,11) SWNT
(the system C$_{60}$@SWNT), Ref.~\cite{sgz06nt}  analyzed the oscillatory motion between the edges of
a finite-length (10,10) SWNT, and Ref.~\cite{hx09pla} demonstrated that the energy dissipation
of the oscillatory motion of C$_{60}$ depends substantially on the radius of SWNT and the presence
of defects. Thermophoretic motion of buckyball enclosed in a series SWNTs was analyzed in Ref.~\cite{rh10cpl}.
It was shown that inside the nanotube the encapsulated fullerenes move in the direction opposing
the thermal gradient, i.e. in the direction of the thermal flow. Similar behavior of the C$_{60}$
molecule was predicted for a graphene sheet where the fullerenes was shown to move
in the direction of a heat flow \cite{lne11pre}.

\begin{figure*}[t]
\begin{center}
\includegraphics[angle=0, width=1\linewidth]{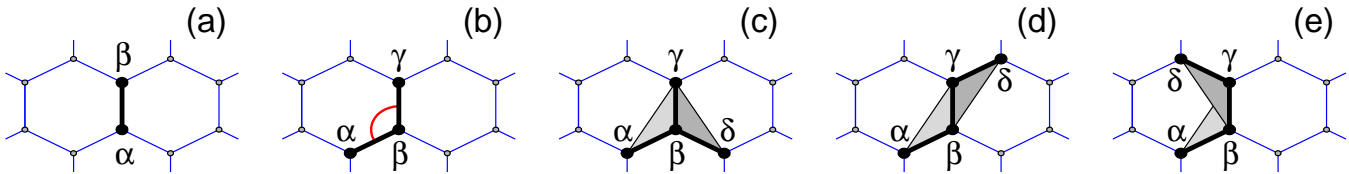}
\end{center}
\caption{\label{fig02}\protect
Typical configurations of the nanoribbon bonds 
containing up to $i$-th nearest-neighbor interactions for:
(a) $i=1$, (b) $i=2$, (c) $i=3$, (d) $i=4$, and (e) $i=5$.
}
\end{figure*}

In this paper, 
we study the dynamics of a molecular system consisting of a fullerene molecule C$_{60}$ or
short-length carbon nanotubes placed on a graphene nanoribbon, as shown in Figs.~\ref{fig01}(a-d).
We employ numerical molecular-dynamics simulations taking into account the motion of all carbon atoms.
Inside of the fullerene molecule and a sheet of graphene the molecules two neighboring carbon atoms
create a valent bond [see Fig. \ref{fig02} (a)]. We describe the energy associated with
the deformation of the valent bond created by two carbon atoms by the following interaction potential
\begin{equation}
U_1(\rho)=\epsilon_1\{\exp[-\alpha (\rho-\rho_1)]-1\}^2,
\label{f2}
\end{equation}
where $\rho$ is the length of the valent bond, $\epsilon_1=4.9632$~eV and $\rho_1=1.418$~\AA~
is the energy and the equilibrium length of the bond, parameter $\alpha=1.7889$~\AA$^{-1}$.
Each carbon atom is placed at the vertex of three planar valent angles.
The corresponding deformation energy of a plane valent angle created by three atoms
[see Fig. \ref{fig02} (b)] can be described by the interaction potential of the form,
\begin{equation}
U_2(\varphi)=\epsilon_2(\cos\varphi+1/2)^2,
\label{f3}
\end{equation}
where $\varphi$ is value of the valent angle (equilibrium value of the angle $\varphi=2\pi/3$),
energy $\epsilon_2=1.3143$~eV. Each valent bond is simultaneously belong to two planes, so
that the deformation energy of the angle formed by two such planes can be described
by the following interaction potential,
\begin{equation}
U_i(\phi)=\epsilon_i(1-\cos\phi),
\label{f4}
\end{equation}
where $\phi$ -- is the corresponding angle (in equilibrium, $\phi=0$), and the index
$i=3$, 4, 5 describe the type of the dihedral angle -- see Fig. \ref{fig02} (c), (d), (e).
Energy $\epsilon_3=\epsilon_4=0.499$~eV, $\epsilon_5\ll \epsilon_4$, so that
the latter contribution to the total energy can be neglected.

More detailed discussion and motivation of our choice of the interaction potentials 
(\ref{f2}), (\ref{f3}), (\ref{f4}) can be found in Ref.~\cite{skh10}. Such potentials have been employed
for modeling of thermal conductivity of carbon nanotubes~\cite{skh09,shk09}
graphene nanoribbons~\cite{skh10} and also in the analysis of their oscillatory modes~\cite{sk09,sk10,sk10prb}.

We describe the nonvalent interaction of carbon atoms belonging to the fullerene molecule
and graphene by the Lennard-Jones 12-6 potential
\begin{equation}
U_{LJ}(r)=4\epsilon_0\left[\left(\frac{\sigma}{r}\right)^{12}-\left(\frac{\sigma}{r}\right)^6\right],
\label{f5}
\end{equation}
where $r$ -- distance between the atom centers, interaction energy $\epsilon_0=\epsilon_{\rm CC}=0.002635$~eV,
and parameter $\sigma=\sigma_{\rm CC}=3.3686$~\AA. Potential (\ref{f5}) coincides with the potential
(\ref{f1}) for the coefficients $C_{12}=4\epsilon_{CC}\sigma_{\rm CC}^{12}=22500$~eV/\AA$^{12}$,
$C_6=4\epsilon_{\rm CC}\sigma_{\rm CC}^6=15.4$~eV/\AA$^6$ (see \cite{umh03prl}).
\begin{figure}[tbp]
\begin{center}
\includegraphics[angle=0, width=1.\linewidth]{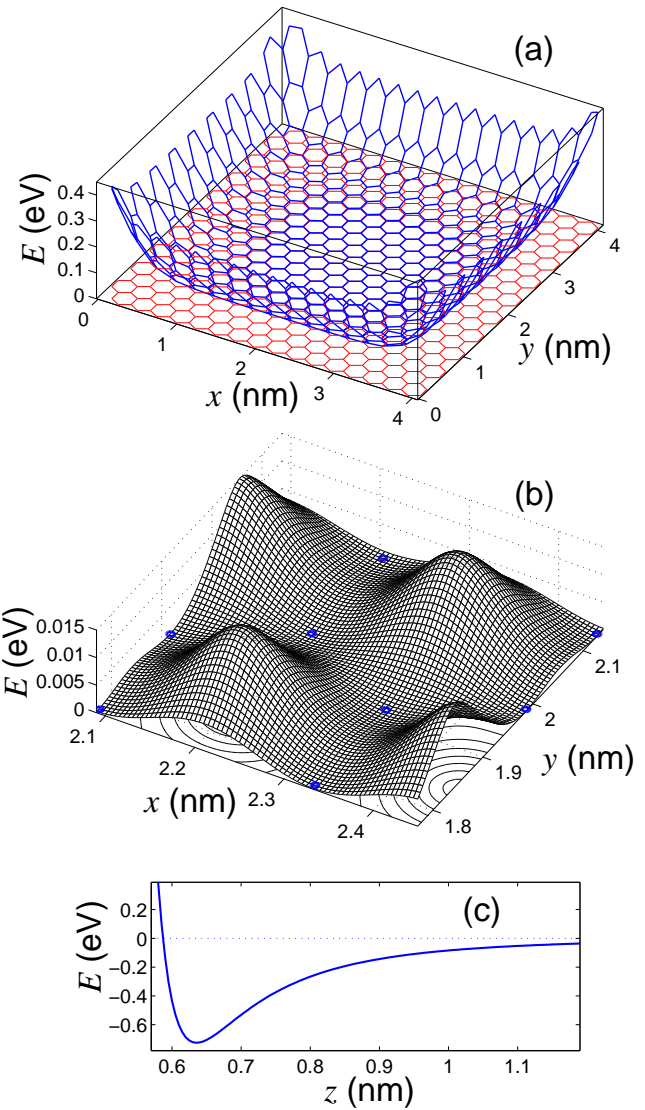}
\end{center}
\caption{\label{fig03}\protect
(a) Energy $E$ of the fullerene C$_{60}$ molecule as a function of
its position on a square sheet of a single-layer graphene of the size $4.1\times 4.1$~nm$^2$.
Red lines mark the valent bonds of the graphene, where the point $(x,y)$ defines the projection
of the molecule center on the plane. (b) Shape of the potential energy $E(x,y)$ at
the center of the graphene sheet, circles mark the position of the carbon atoms of the
graphene. (c) Interaction energy as a function of the distance $z$ between the center
of C$_{60}$ molecule and the plane of graphene.
}
\end{figure}
\begin{figure}[tb]
\begin{center}
\includegraphics[angle=0, width=1\linewidth]{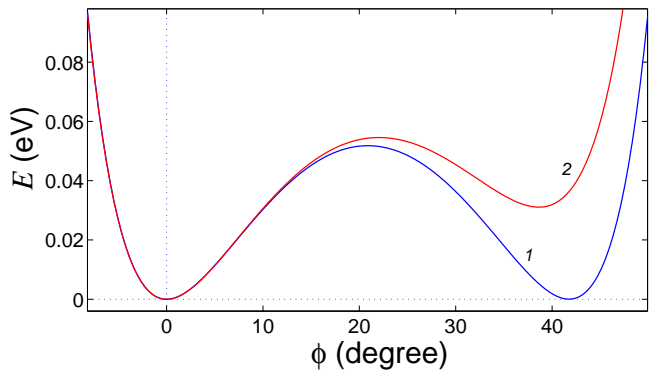}
\end{center}
\caption{\label{fig04}\protect
Change the energy of C$_{60}$ molecule due its rotation around the edge closest to the
graphene sheet. Curve 1: rotation from one hexagon face to the neighboring hexagon face;
curve 2: rotation from the hexagon face to the neighboring pentagon face.
}
\end{figure}

\section{Approximation of rigid molecules}

First, we study the energy of the coupled system "C$_{60}$ + nanoribbon" as a function of the
relative position of the buckyball on the surface. We consider a square sheet of graphene
of the size $4.1\times 4.1$~nm$^2$ [see Fig.~\ref{fig03}~(a)] and assume that both the C$_{60}$
molecule and the graphene are in the ground state and all valent bonds are not stretched,
and the angles between the bonds do not deform. Then the energy of the complex is defined
by a sum of all nonvalent interactions (\ref{f5}) between the carbon atoms
of the fullerene molecule and the graphene substrate.

Without restrictions of generality, we assume that the sheet of graphene remains at rest in the $(x,y)$
plane, so that the position of the C$_{60}$ molecule is defined by the coordinate  $(x,y)$ of its
center on the plane and the distance $z$ from the center to the substrate surface. We fix the center
coordinate  $(x,y)$ and minimize the energy $E$ of the whole molecular complex as a function of
the distance $z$ and all angles that define the orientation of the C$_{60}$ molecule in space.
Then we obtain two-dimensional function $E(x,y)$ that describes a change of the energy of the
system when the fullerene molecule changes its position on the graphene sheet.

The effective potential energy $E(x,y)$ shows that in the ground state the C$_{60}$ molecule takes the
position above the center atom of the graphene sheet at the distance $z_0=0.635$~nm  from the surface,
so that the hexagonal face closest to the surface is parallel to it. Figure~\ref{fig03}~(c) shows
the dependence of the energy $E$ on the distance $z$ to the surface. As follows from this dependence,
in order to escape the surface the molecule requires the energy $\Delta E_0=0.73$~eV.
In the earlier study~\cite{narne10pre}, similar values were defined as $z_0=0.65$~nm and
$\Delta E_0=0.76$~eV, and the experimental evaluation of the coupling energy \cite{umh03prl}
gave the value $\Delta E_0=0.85$~eV.

Dependence of the energy $E$ of the coupled molecular systems on the position coordinate $(x,y)$
of the molecule center on the plane of the graphene sheet is shown in Figs.~\ref{fig03}~(a,b).
At the center of the sheet, the energy depends on the position of the fullerene molecule, and the
positions above the carbon atoms of the substrate are more favorable and they correspond to local minima.
The energy profile suggests that the molecule may slide along the graphene sheet with its hexagonal
face parallel to the surface following the structure defined by the valent bonds of the substrate.
This motion would require overtaking the energy barrier $\Delta E_1=0.0015$~eV, whereas the
sliding motion in other directions would require higher energy to overtake the barrier
$\Delta E_2=0.015$~eV. We notice that the sliding of the C$_{60}$ molecule between two parallel
sheets of graphene was studied earlier in Ref.~\cite{iams10jp},
where was found that the corresponding barriers are: $\Delta E_1=0.002$~eV and $\Delta E_2=0.02$~eV.

When the molecule approaches the edge of the nanoribbon, the total energy of the system  $E(x,y)$
grows, as shown in Fig.~\ref{fig03}~(a), and at the edge of the nanoribbon potential barrier  
$\Delta E_3>0.3$~eV repels the molecule from the edge.

In addition to the sliding, the C$_{60}$ molecule may also roll along the surface. To evaluate when
this rolling becomes possible, we analyze a change of the energy of the ground state when
the molecule rotates around the edge of its hexagonal face closest to the graphene surface.

In the ground state the hexagonal face of the fullerene closest to the graphene sheet is parallel
to the substrate surface. If we fix one edge of this face and rotate the buckyball around it
modeling the rocking of the C$_{60}$ molecule, we can find the dependence of the energy $E$ on the
rotation angle $\phi$, shown in Fig.~\ref{fig04}. For such a procedure, two cases are possible.
In the first case, the molecule rotates replacing one hexagonal face by another similar face,
the energy $E(\phi)$ for this rotation has the form of a symmetric two-well potential, as shown
in Fig.~\ref{fig04} (curve 1). For the other type of rotation, the hexagonal face is replaced by a pentagonal
face also parallel to the substrate surface, in this later case the energy has the form of
asymmetric two-well potential also shown in Fig.~\ref{fig04} (curve 2).
In both these cases of the molecule rolling, the molecule should overtake the energy barrier
$\Delta E_4=0.052$~eV substantially higher than the energy barrier $\Delta E_2$ required for its sliding.

Our analysis suggests that the  C$_{60}$ molecule may create long-lived coupled states with the
graphene nanoribbons or graphene sheets of a finite extent, and its sliding along the surface is
more energetically favorable than its rocking or rolling. Thus, we may expect that for low
temperatures the main type of motion of the fullerene molecules would be sliding, and it will be
replaced by rolling at higher temperatures.
To verify these ideas, below we provide the numerical modeling of the motion of the fullerene molecules
along the graphene surface taking into account all degrees of freedom of the coupled system
and considering different values of temperature.
\begin{figure}[tb]
\begin{center}
\includegraphics[angle=0, width=1\linewidth]{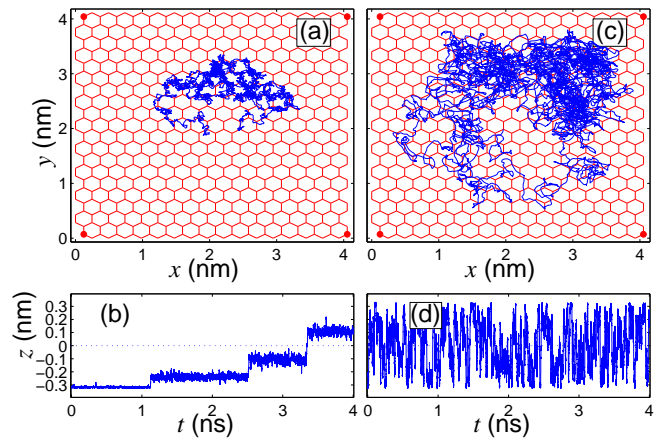}
\end{center}
\caption{\label{fig05}\protect
Stochastic dynamics of the fullerene molecule C$_{60}$ placed on a graphene sheet
with the size of $4.1\times 4.1$~nm$^2$. (a,c) Trajectory of the molecule center shown as a projection
on the plane of graphene at the temperatures $T=30$K and 300K, respectively. Red color shows the
valent bonds of carbon atoms of the graphene, red dots show the fixed atoms.  (b,d) Temporal
evolution of the $z$ projection of the center of the hexagonal face relative the molecule's
center of mass, for  $T=30$K and 300K, respectively.
}
\end{figure}

\section{Stochastic motion of $C_{60}$ on a graphene sheet}

First, we study the stochastic motion of the fullerene molecule $C_{60}$ placed on a thermalize
sheet of graphene. We consider a flat square sheet of graphene of the size $4.1\times4.1$~nm$^2$
composed of 678 carbon atoms in the $(x, y)$ plane, as shown in Figs.~\ref{fig05}(a,c). To keep
the graphene in the plane, we fix the four edge atoms, shown by solid red circles in the corners 
of the plots of Fig. (a,c).
We place the molecule C$_{60}$ at the middle of the sheet and place the whole system
"buckyball + graphene" \ consisting of $N=678+60$ atoms into Langevin thermostat modeled
by stochastic forces. The motion equations are taken in the form
\begin{equation}
M_i\ddot{\bf u}_i=-\frac{\partial H}{\partial {\bf u}_i}-\Gamma M_i\dot{\bf u}_i+\Xi_i,
\label{f6}
\end{equation}
where $H$ is Hamiltonian of the molecular complex, the indices $i=1,2,...,N$ stand for the atom number,
the vector ${\bf u}_i=(u_{i,1},u_{i,2},u_{i,3})$ describes the position of the atom, and $M_i$ is
the atom mass. Langevin collision frequency is $\Gamma=1/t_r$, where $t_r$ -- particle relaxation time,
and $\Xi_i=(\xi_{i,1},\xi_{i,2},\xi_{i,3})$ is a three-dimensional vector of Gaussian stochastic forces
describing the interaction of the $i$-th atom with thermostat.

We consider a hydrogen-terminated graphene, where all edge atoms correspond
to the CH group. In our model, we treat such a group as a single effective particle at the
location of the carbon atom, so that we take the mass of atoms inside the graphene sheet and
fullerene atoms as $M_i=12m_p$, but for the atoms at the graphene edge we consider an effective
particle with a larger mass $M_i=13m_p$ (where $m_p=1.6603\times 10^{-27}$ kg is the proton mass).

To analyze the thermalized dynamics of the system, we introduce a random white noise with the
correlation functions
\begin{eqnarray}
\langle\xi_{i,k}(t_1)\xi_{j,l}(t_2)\rangle=2M_i\Gamma k_BT\delta_{ij}\delta_{kl}\delta(t_2-t_1),\\
\label{f7}
i,j=1,2,...,N,~~k,l=1,2,3, \nonumber
\end{eqnarray}
where  $T$ is the temperature of the Langevin thermostat. We select relatively large relaxation
time $t_r=1$~ps to reduce the effect of viscosity in the molecular dynamics, and integrate
the system of equations (\ref{f6}) for the time interval $t=4$~ns (for the period $t=10t_r=10$~ps
the system should relax into the equilibrium state with the thermostat and then it will evolve
as a fully thermalized system)

Our numerical simulations demonstrate that for the temperatures $T\le 500$K the molecular system
preserves its geometry, namely the buckyball C$_{60}$ remains attached to the graphene sheet, and
it undergoes stochastic motion not approaching the edge of the sheet. For very low temperatures
$T<150$K the main type of motion is sliding along the surface, while for higher temperatures
the sliding is replaced by rolling.
Figures~\ref{fig05}(a,b) show an example of such a motion at $T=30$K when the buckyball moves
along the valent bonds of the graphene sheet [see Fig.~\ref{fig05} (a)] changing time to time its
sliding face [see Fig.~\ref{fig05} (b)], such that for the time period of 4 ns it changed the
face 3 times, as shown in  Fig.~\ref{fig05}(b). For higher temperatures such as $T=300$K the dynamics
is clearly stochastic, and the motion does not reflect much the discrete structure of the graphene
substrate, with the contacting face being changed continuously [see Figs.~\ref{fig05}(c,d)].
We observe that in this case the main type of motion is rolling along the graphene sheet,
which is energetically more favorable than sliding.
\begin{figure}[tb]
\begin{center}
\includegraphics[angle=0, width=1\linewidth]{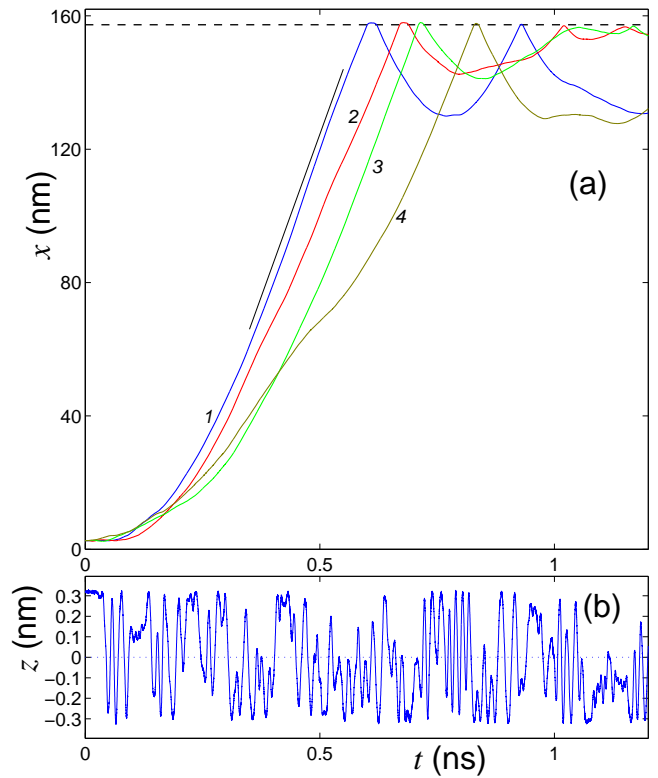}
\end{center}
\caption{\label{fig06}\protect
Transport of the molecular complex Fe$^-$@C$_{60}$ along the nanoribbon  with the dimensions
$157\times 2$~nm$^2$, under the action of the constant electric field $E=10^{-7}$ V/m, at temperature
$T=300$K. (a) Evolution of the coordinate of the molecule center for four different thermalizations
of the nanoribbon (curves 1,2,3, and 4). An auxiliary straight line show the slope of the curves
corresponding to the maximum velocity of the molecule, $v_m=390$~m/s. Dashed straight line marks
the location of the edge of the nanoribbon. (b) Temporal evolution of the $z$ projection of
the center of the hexagonal face relative the molecule's center of mass, corresponding
to the curve 1 in (a).
}
\end{figure}

\section{Electorphoresis of Fe$^-$@C$_{60}$}

In order to model numerically the motion of a spherical C$_{60}$ molecule along the graphene
nanoribbon under the action of a constant electric field, we include an ion of iron Fe$^-$
into the buckyball, and consider a composite structure Fe$^-$@C$_{60}$.
\begin{figure}[t]
\begin{center}
\includegraphics[angle=0, width=1\linewidth]{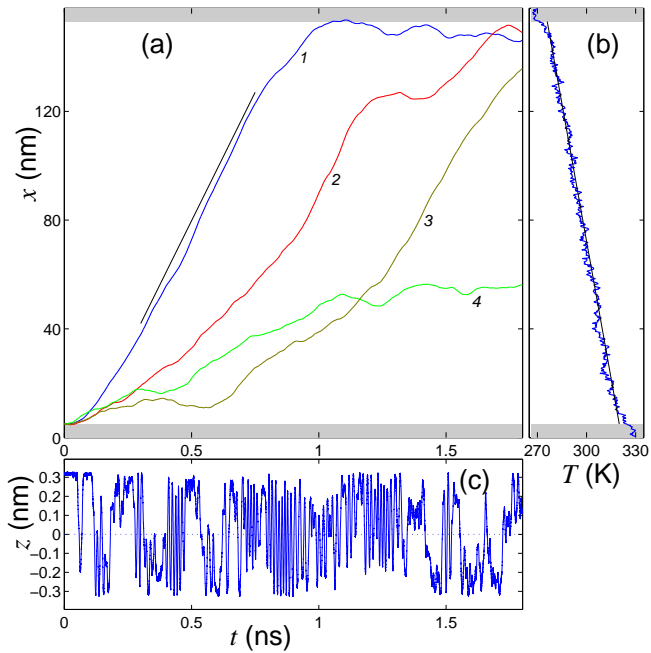}
\end{center}
\caption{\label{fig07}\protect
Transport of the C$_{60}$ molecule along the nanoribbon with the size $157\times 2$~nm$^2$,
under the action of a heat flow. (a) Evolution of the coordinate of the molecule center
for four different thermalizations of the nanoribbon (curves 1,2,3, and 4). An auxiliary straight
line shows the slope of the curves corresponding to the maximum velocity of the molecule, $v_m=190$~m/s.
(b) Temperature profile along the nanoribbon, a solid straight line corresponds to the temperature
gradient of $dT=0.30$~K/nm. Grey shaded areas show the edge areas of the nanoribbon subjected
to two Langevin thermostats at $T_+=330$~K and $T_-=270$~K, respectively. (c) Temporal evolution
of the $z$ projection of the center of the hexagonal face relative the molecule's center of mass,
corresponding to the curve 1 in (a).
}
\end{figure}

We describe the interaction of the iron molecule with the carbon atoms of the molecule C$_{60}$
by employing the LJ potential (\ref{f5}) with the interaction energy $\epsilon_0=\epsilon_{\rm FeC}=0.0031$~eV
and $\sigma=r_{\rm FeC}/2^{1/6}$, where the equilibrium distance between the atoms
is $r_{\rm FeC}=3.51$~\AA. We study the motion of the molecular complex Fe$^-$@C$_{60}$ along the
zigzag nanoribbon consisting of 640 segments with 20 carbon atoms in each. Such a ribbon has the
dimensions $157\times 2$~nm$^2$ and it consists of 12800 carbon atoms (a part of the nanoribbon is
shown in Figs.~\ref{fig01} (a-d).

We place an undeformed nanoribbon in the $(x,y)$ plane along the axes $x$ and fix
it in the plane by pinning its four edge atoms. Then, we place the system into the Langevin thermostat
by solving numerically the system of stochastic equations (\ref{f6}) with $\Gamma=1/t_r$, where $t_r=1$~ps.
During the time $t=10t_r$ the system will get thermalized and then we add the molecular complex Fe$^-$@C$_{60}$.
By applying the constant electric field $E$ along the axis $x$, we create a constant external
force ${\bf F}=(1,0,0)eE$, where $e$ is the electron charge.

Typical dynamics of Fe$^-$@C$_{60}$ molecule is presented in Fig.~\ref{fig06}. Under the
action of the electric field the composite molecule moves along the nanoribbon towards its right
end not detaching from it. The speed of motion grows monotonously before reaching the maximum 
velocity $v_m$.
For the field $E=10^{-7}$~V/m, the maximum velocity $v_m=390$~m/s which is not sufficient to
overtake the effective potential barrier created by the nanoribbon edge, so the molecular complex
is reflected from the right end and remains on the nanoribbon, as seen from Fig.~\ref{fig06}~(a).
However, for higher strength of the external field $E=2\times 10^{-7}$ V/m the molecule reaches
the maximum velocity $v_m=540$~m/s, and it may drop off the right end of the nanoribbon.
More detailed analysis indicates that the sliding motion is observed only at the beginning of the motion,
whereas the main motion of the Fe$^-$@C$_{60}$ molecule is its rolling along the nanoribbon
in the direction of the applied electric field, see Fig.~\ref{fig06}~(b).
\begin{figure}[t]
\begin{center}
\includegraphics[angle=0, width=1\linewidth]{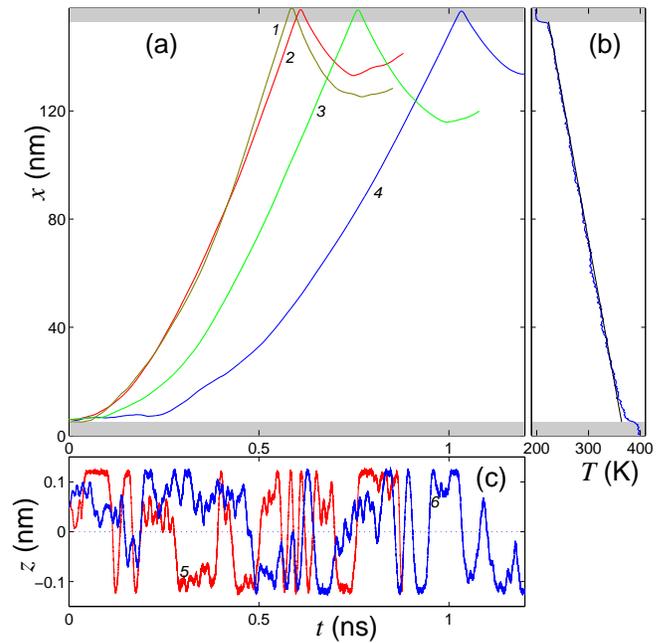}
\end{center}
\caption{\label{fig08}\protect
(a) Transport of the spherical C$_{60}$ molecule, and three different short-length (5,5) carbon nanotubes
C$_{100}$, C$_{180}$ and C$_{260}$  along the zigzag nanoribbon of the size
$157\times 2$~nm$^2$ under the action of a heat flow. Shown is the evolution of the coordinate
of the molecule center for four cases (curves 1, 2, 3 and 4, respectively).
(b) Temperature profile along the nanoribbon, a solid straight line corresponds to the temperature
gradient of $dT=0.95$~K/nm.  Grey shaded areas show the edge areas of the nanoribbon subjected
to two Langevin thermostats at $T_+=400$~K and $T_-=200$~K, respectively. (c) Temporal evolution
of the $z$ projection of the center of the pentagonal face relative the molecule's center of mass,
corresponding to the curves 2 and 4 in (a).
}
\end{figure}

\section{Thermophoresis of C$_{60}$ molecules}

In order to model the motion of a spherical C$_{60}$  
molecule under the action of a constant heat flow, we
consider the graphene nanoribbon of the size $157\times 2$~nm$^2$ and place its edges into the
Langevin thermostats at different temperatures: first 20 sections -- at the temperature
$T_+=330$~K, and the last 20 sections, at the temperature $T_-=270$~K. After some transition period,
we observe the formation of a constant temperature gradient $dT=0.30$~K/nm, as shown in
Fig.~\ref{fig07}~(b). Then, we place a molecule C$_{60}$ on the 22-nd section of the nanoribbon
and study its temperature-driven dynamics.

Our modeling demonstrates that C$_{60}$ molecule moves towards the right end of the nanoribbon
under the action of the temperature gradient with the constant velocity $v_m=190$~m/s, see Fig.~\ref{fig07}~(a).
At the initial stage, the typical motion of the molecule is sliding, but after
some time the molecule starts rolling along the nanoribbon, see Fig.~\ref{fig07}~(c).

For larger temperature difference, e.g. $T_+=400$~K and $T_-=200$~K, the temperature gradient is
larger, $dT=0.95$~K/nm [see Fig.~\ref{fig08}~(b)], and the molecule speeds up reaching the maximum velocity
$v_m=440$~m/s, and then it, is reflected elastically from the edge of the nanoribbon, see Fig.~\ref{fig08}~(a).

\section{Thermophoresis of carbon nanotubes}

To compare the transport properties of fullerenes and carbon nanotubes and evaluate the effect of
shape on thermophoresis, we consider the mobility of short-length  (5,5) carbon nanotubes
C$_{100}$, C$_{180}$ and C$_{260}$. These nanotubes have the form of short cylinders with the diameter
$D=0.70$~nm (that equals to the diameter of the C$_{60}$  molecule) but with different lengths
$L=1.18$, 2.19 and 3.17~nm, respectively [cf. Fig.~\ref{fig01}(a) and Figs.~\ref{fig01}(b-d)].
The interaction energy will be maximal provided the nanotube is place along the nanoribbon,
so that rocking is hardly possible because for this latter case 
the nanotube should be placed across the
nanoribbon. Therefore, we anticipate that the most typical motion for short-length nanotubes would be
their sliding along the surface of the nanoribbon. This is indeed confirmed by our numerical modeling.

Again, we consider the graphene nanoribbon of the size $157\times 2$~nm$^2$ and place its edges
into the Langevin thermostats at different temperatures: first 20 sections -- at the
temperature $T_+=400$~K, and the last 20 sections, at the temperature $T_-=200$~K.
After some transition period, we observe the formation of a constant temperature gradient
$dT=0.95$~K/nm, as shown in Fig.~\ref{fig08}~(b). Then, we place a nanotube next to the
left end of the nanoribbon and study its temperature-driven dynamics.

Our modeling demonstrates that the nanotube moves towards the right end of the nanoribbon under
the action of the temperature gradient with the constant velocity $v_m$, see Fig.~\ref{fig08}~(a).
The maximum velocities are: $v_m=444$~m/s, for  $C_{60}$ molecule, and $v_m=396$, 328 and 283~m/s,
for the nanotubes  C$_{100}$,  C$_{180}$, and C$_{260}$, respectively. Moreover, our study 
indicates that the nanotubes do not roll but instead they slide along the surface, this is indicated
by the study of the temporal evolution of the $z$ projection of the center of the pentagonal face
relative the nanotube's center of mass, as shown in Fig.~\ref{fig08}~(c).

\section{Conclusions}

We have studied the motion of fullerene molecules and short-length carbon nanotubes along the
deformable surface of a graphene nanoribbon in several cases, including the stochastic motion and
directed motion in the presence of an electric field or a thermal gradient. We have revealed that
the mobility of carbon structures depends substantially on their shape and symmetry, so that
larger molecules and more asymmetric structures demonstrate much lower mobility than $C_{60}$ molecules.
We have demonstrated that both electorphoresis and thermophoresis can be employed to control carbon
structures and fullerenes through their shape, size, and possible inclusions.

\section{Acknowledgements}

Alex Savin acknowledges a warm hospitality of the Nonlinear Physics Center at the Australian
National University, and he thanks the Joint Supercomputer Center of the Russian Academy 
of Sciences for the use of their computer facilities. 
The work was supported by the Australian Research Council.

\end{document}